\newcommand{\jpsi}      {\mbox{$J/\psi$}}
\newcommand{\gev}{{\rm Ge}\kern-1.pt{\rm V}}
\newcommand {\pom} {I\hspace{-0.2em}P}

\documentclass[proceedings]{JHEP3}

\PrHEP{PrHEP hep2001}                   
\conference{International Europhysics Conference on HEP}                

\usepackage{epsfig}

\title{Elastic Vector Meson Production at HERA}

\author{\speaker{Bruce Mellado}  (On behalf of the H1 and ZEUS Collaborations)   \\
        Columbia University, 538 120th Street West, New York, NY 10027, USA                                         
        E-mail: \email{mellado@nevis1.nevis.columbia.edu}}

\abstract{The H1 and ZEUS Collaborations report  new results on
elastic VM cross-section and trajectory determination. Elastic VM
production appears to be independent of the photon polarization.
The presence of non-zero shrinkage in the photoproduction of
$\jpsi$ indicates the presence of soft physics. Accumulated data
do not accommodate a universal Pomeron trajectory. The steepness
of the $x$ dependence of elastic VM production with changing
$Q^2+M_V^2$ is similar to that of $F_2$ with changing $Q^2$.}

\begin{document}

  \section{Introduction}

The production of vector mesons (VM, $V$) at HERA is interesting
for the study of non-perturbative hadronic physics, perturbative
QCD (pQCD) and their interplay, and the sensitivity to saturation
effects. Moreover, VM production is complementary to
deep-inelastic scattering (DIS). Results from DIS have shown the
correctness of pQCD down to low values of $Q^2\sim$ 1 GeV$^2$
(where $Q^2$ is the virtuality of the exchanged photon), a region
where a transition to non-perturbative physics is
observed~\cite{epj_7_609}. In the past, the elastic processes were
mostly treated through non-perturbative methods. These methods
successfully describe the basic features of exclusive light VM
production at low photon virtualities. However, in recent years a
pQCD picture of the exclusive VM production is able to describe
the basic features provided that $Q^2/\Lambda^2_{QCD}\gg$ 1 and
$M_{V}/W\ll$ 1, where $M_{V}$ is the mass of the VM and  $W$ is
the photon-proton center-of-mass energy. The photon fluctuates
into a quark-antiquark pair, $q\overline{q}$, a color dipole, long
before the interaction with the proton occurs. The interaction
occurs via gluon ladders. The steepness of the rise of the VM
production with $W$ is basically driven by the gluon density in
the proton, which is probed at an effective scale, $\mu$,
$\mu^2\sim(Q^2+M^2_{V})$ at low $x$ ($x\simeq(Q^2+M_V^2)/W^2$).

Assuming a flavor independent production mechanism, the relative
production rates should scale approximately with the square of the
quark charges, i.e. the relative production rates scale as
$\rho:\omega:\phi:J/\psi=9:1:2:8$, referred to here as SU(4)
ratios. It is interesting to determine how the $q\overline{q}-p$
scattering changes the SU(4) ratios.

As of today the H1 and ZEUS Collaborations~\footnote{A compilation
of references may be found in~\cite{hep-ph_00_11050}. In addition,
new contributions from the H1 and ZEUS collaborations are reviewed
in the present contribution.} have measured the elastic production
of VM's, $e p\rightarrow e V p$, where $V=\rho,\omega,\phi$ and
$J/\psi$ over a wide range of $W$ and  from photoproduction
($Q^2\simeq 0$) to $Q^2=100$ GeV$^2.$
The cross-section for ${q\overline{q}}$ scattering on the proton,
$\sigma_{q\overline{q}}$, contains the dynamics with which the
elastic VM production and total DIS cross-sections may be
described. We are in a position to investigate the $x$ (or $W$)
dependence of the VM production cross-section ordered according
$Q^2+M^2_V$ and to investigate similarities between this and the
$x$ dependence of the DIS total cross-section.




\section{Results}

The H1 Collaboration reports new results on the elastic
photoproduction of $\rho$ mesons, where the high momentum proton
is tagged in the forward proton spectrometer
(FPS)~\cite{eps01_810}. This experimental technique has the
advantage that the elastic reaction is tagged directly. No
subtraction of the proton dissociative production is needed.
Additionally, by applying this method, one is able to measure
directly the $t$ dependence of the cross-section, where $t$ is the
square of the four-momentum transfer at the proton vertex. The
kinematic range of the measurement is $25<W<70\,\gev$,
$0.073<\left|t\right|<0.4\,\gev^2$ and
$E_{p^{\prime}}/E_{p}>0.98$, where $E_{p}$ and $E_{p^{\prime}}$
are the energy of the incoming and scattered proton, respectively.


\FIGURE[htb]{\epsfig{figure=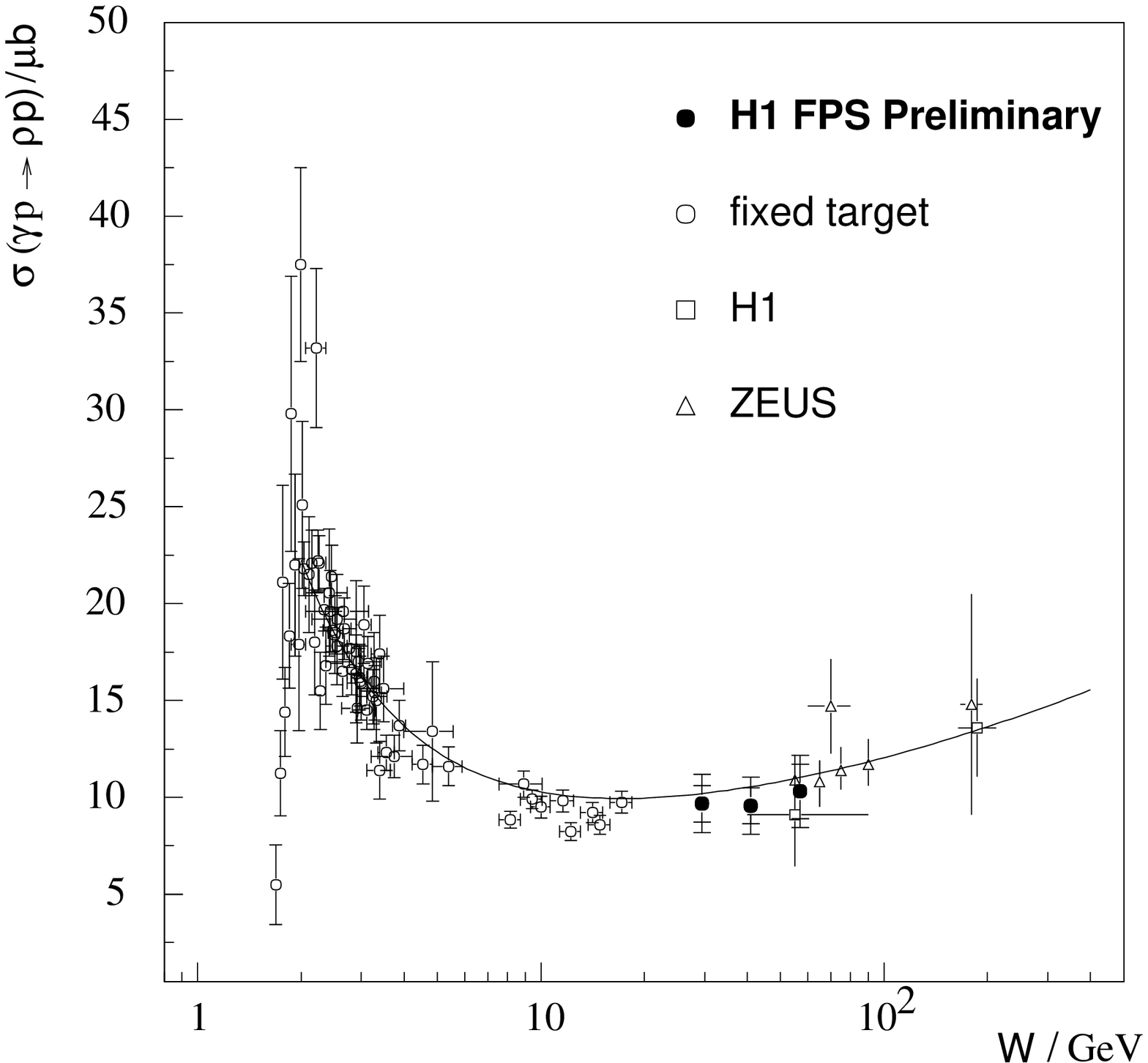,width=6.1cm}
\caption{The elastic $\rho$ meson production cross-section
$\sigma(\gamma p\rightarrow \rho p)$ compared with other
measurements. The solid line corresponds to a fit by DL (see
text).} \label{fig:fps}}

 The elastic cross-section is measured in three bins of $W.$ The
results are shown in Figure~\ref{fig:fps} compared with other
measurements performed at HERA and fixed target experiments,
closing the gap in $W$ between them. The solid line corresponds to
a fit performed on the energy dependence of the elastic scattering
cross-sections in hadron-hadron collisions  by Donnachie and
Landshoff (DL) \cite{pl_296_227}. Also, the slope of the
differential cross-section, $b$, for $d\sigma/dt\propto \exp(bt)$
is measured as a function of $W$. The $t$ dependence of the
Pomeron trajectory
($\alpha_{\pom}(t)=\alpha_{\pom}(0)+\alpha_{\pom}^{\prime}t$,
where $\alpha_{\pom}^{\prime}$ is also referred to as shrinkage)
is compatible with the soft Pomeron \cite{pl_296_227} extracted by
DL.



The ZEUS Collaboration reports new results on the
electroproduction of $\rho$ mesons in the range $32<W<160\,\gev$
and for photon virtualities of
$2<Q^2<80\,\gev^2$~\cite{eps01_594}. The ratio of the production
cross-section, $R=\sigma_L/\sigma_R$, for longitudinally
($\sigma_L$) and transversely ($\sigma_T$) polarized virtual
photons is determined. The ratio increases with $Q^2$; however, it
is constant with $W$. The Pomeron trajectory is extracted for the
first time in the electroproduction regime from VM production (see
Figure~\ref{fig:rhotrajectory}). The trajectory parameters
obtained in the range $2<Q^2<40\,\gev^2$ are
$\alpha_{\pom}(0)=1.14\pm 0.01^{+0.03}_{-0.03}$ and
$\alpha_{\pom}^{\prime}=0.04\pm 0.07^{+0.13}_{-0.04}\,\gev^{-2}.$



\DOUBLEFIGURE[htb]{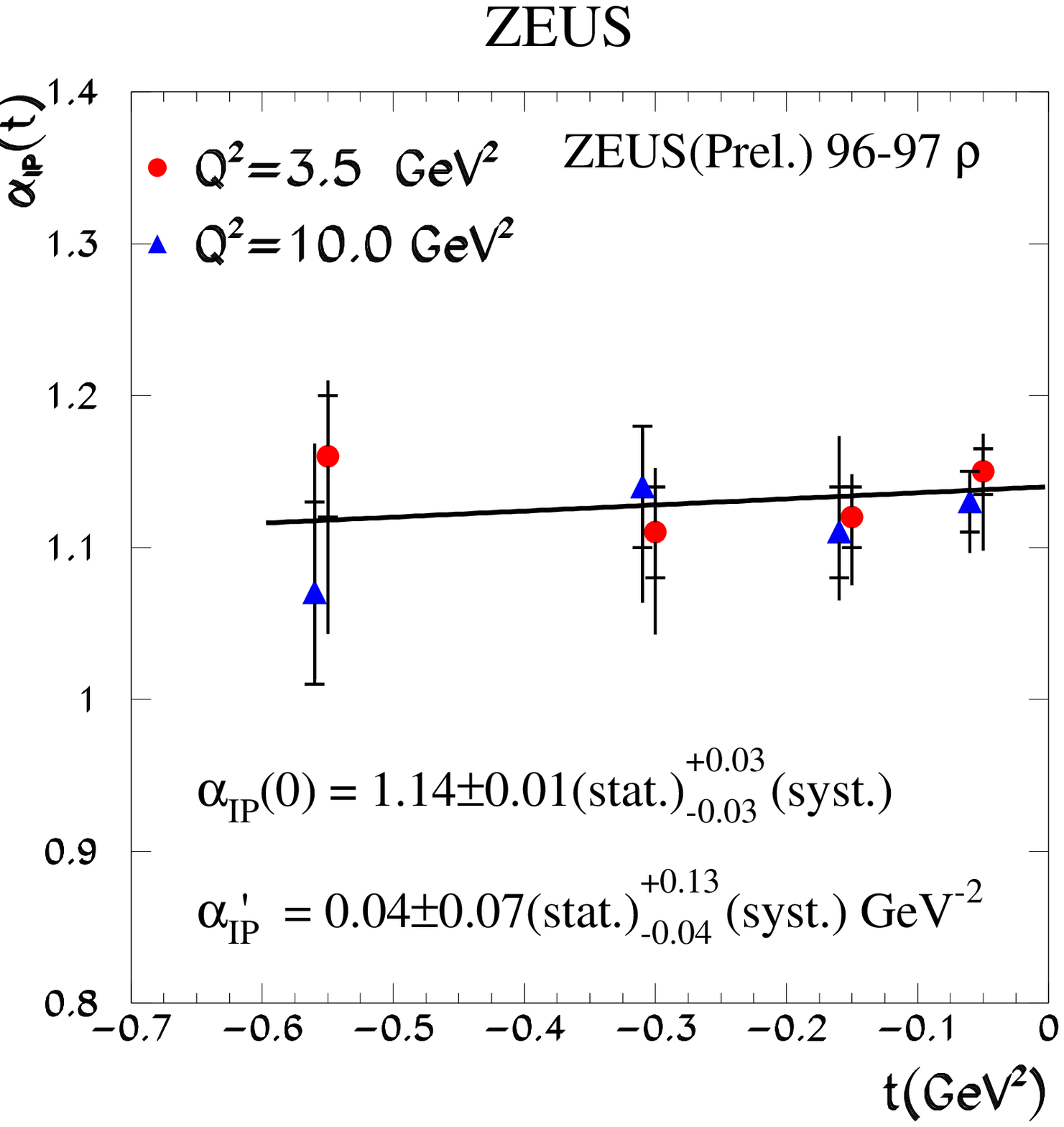,width=6.5cm}{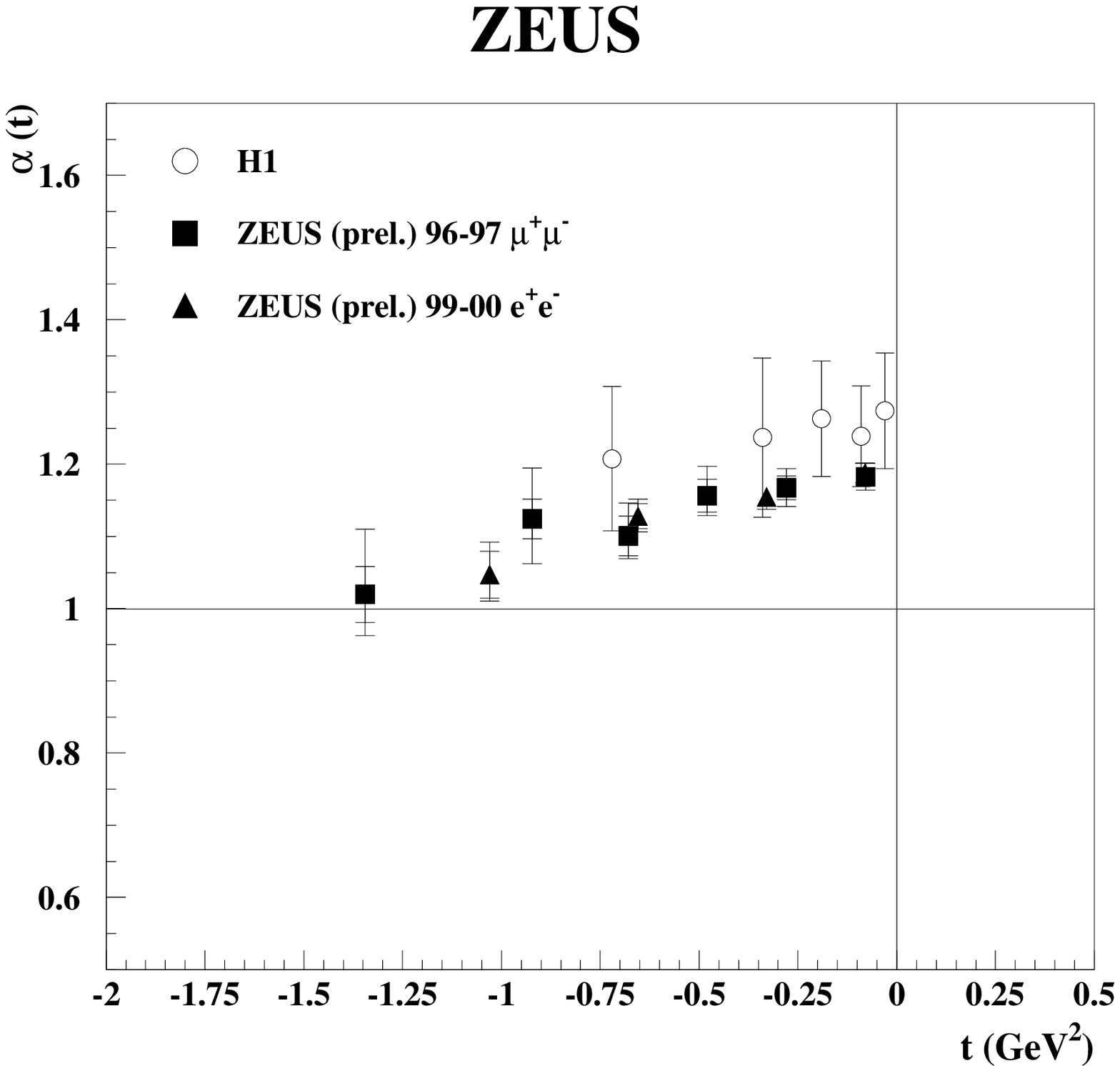,width=6.5cm}{The
Pomeron trajectory from the electroproduction of elastic $\rho$ as
a function of $t$ for two $Q^2$ bins, $2<Q^2<6\,\gev$ and
$6<Q^2<40\,\gev^2$. The fit of the form
$\alpha_{\pom}(t)=\alpha_{\pom}(0)+\alpha_{\pom}^{\prime}t$ is
performed to all 8 points. The results of the fit are shown.
\label{fig:rhotrajectory}}{The Pomeron trajectory from the
photoproduction of elastic $\jpsi$ as a function of $t$ from the
H1~\cite{pl_483_23} and ZEUS experiments. The inner error bars
indicate the  statistical errors, the outer bars are the
statistical and systematic uncertainties added in quadrature.
\label{fig:alpha_comparison}}

The ZEUS Collaboration reports new results on the differential
cross-section, $d\sigma/dt$, of the photoproduction of $\jpsi$ in
an extended range energy range $20<W<290\,\gev$ and
$\left|t\right|<1.25\,\gev^2$, with data corresponding to
$55.3\,{\rm pb}^{-1}$ of luminosity, using the electron decay
channel~\cite{eps01_548}. The $t$-slope has been extracted in bins
of $W$, yielding $b=4.3\pm0.08^{+0.16}_{-0.41}\,\gev^{-2}$ at
$W=90\,\gev$. The Pomeron trajectory has been extracted yielding
$\alpha_{\pom}(0)=1.201\pm 0.013^{+0.003}_{-0.011}$ and
$\alpha_{\pom}^{\prime}=0.126\pm
0.029^{+0.015}_{-0.028}\,\gev^{-2}.$ The value of the latter
parameter indicates the presence of small, but not negligible,
shrinkage. This is illustrated in
Figure~\ref{fig:alpha_comparison}.

The H1 Collaboration presents results of the $t$-slope of the
differential cross-section, $d\sigma/dt$, of the photoproduction
of $\psi(2S)$, studied through the leptonic decay
channels~\cite{ichep00_987}. The value of $b_{\psi(2S)}=4.5\pm
1.2^{+1.4}_{-0.7}\,\gev^2$ is similar to that obtained in the
photoproduction of $\jpsi$. The ZEUS Collaboration reports new
results of the photoproduction cross-section of $\psi(2S)$ studied
through the electron decay channel in the range
$50<W<125\,\gev$~\cite{eps01_562}. The cross-section ratio
$\sigma_{\gamma p\rightarrow \psi(2S) p}/\sigma_{\gamma
p\rightarrow J/\psi p}$ has been calculated. This result is
consistent with previous measurements. Both results by the H1 and
ZEUS Collaborations are consistent with the expectations of the
quark parton model.

\section{Discussion}

The value of $\delta_{J/\psi}\approx 0.7$ for $(\sigma(\gamma p
\rightarrow V p) \propto W^{\delta_V})$ indicates a strong rise
with $W$ for the cross-section of the photoproduction of $\jpsi.$
This behavior is qualitatively different from that observed in the
photoproduction of the $\rho$ meson, with the value of
$\delta_{\rho}\approx 0.16$~\cite{np_463_3,epj_2_247}. This small
value indicates that the production of $\rho$ mesons is dominated
by soft physics. It is expected that the scattering of a
transversely polarized $q\overline{q}$ state on proton is driven
by a soft interaction. In the photoproduction regime, where the
mean value of $Q^2$, $\langle Q^2\rangle\approx 5\times
10^{-5}\,\gev^2\ll M_{J/\psi}^2$, it is expected that, the
$q\overline{q}$ state is transversely polarized~\footnote{The
ratio of longitudinal to transversely polarized production of VM's
is proportional to $Q^2/M_V^2$ for $Q^2\lesssim 1\,\gev^2.$}.
Additionally, it has been observed that the value of
$\delta_{J/\psi}$ does not increase with $Q^2$ in the
electroproduction regime~\cite{ichep00_879}.  This indicates that
the $W$ dependence cross-section for longitudinally and
transversely polarized $\jpsi$ are similar for the same final
state. This is consistent with the observation that
$R=\sigma_L/\sigma_T$ in the electroproduction of $\rho$ mesons is
constant with $W$~\cite{ichep00_880}. This leads to the conclusion
that, despite the large differences in the dipole size expected
for longitudinally and transversely polarized photons, the elastic
production of VM's appears to be independent of the photon
polarization.

Strong shrinkage is indicative of soft physics. In processes
dominated by hard physics it is expected that
$\alpha_{\pom}^{\prime}$ is zero or very small. However, it has
been observed that the value of $\alpha^{\prime}_{\pom}$ obtained
from the photoproduction of $\jpsi$ deviates from zero by some
four standard deviations. It is not clear whether pQCD can
accommodate this large value. This suggests that the interaction
leading to the photoproduction of $\jpsi$ has a sizeable component
of soft physics.

\FIGURE[t]{\epsfig{figure=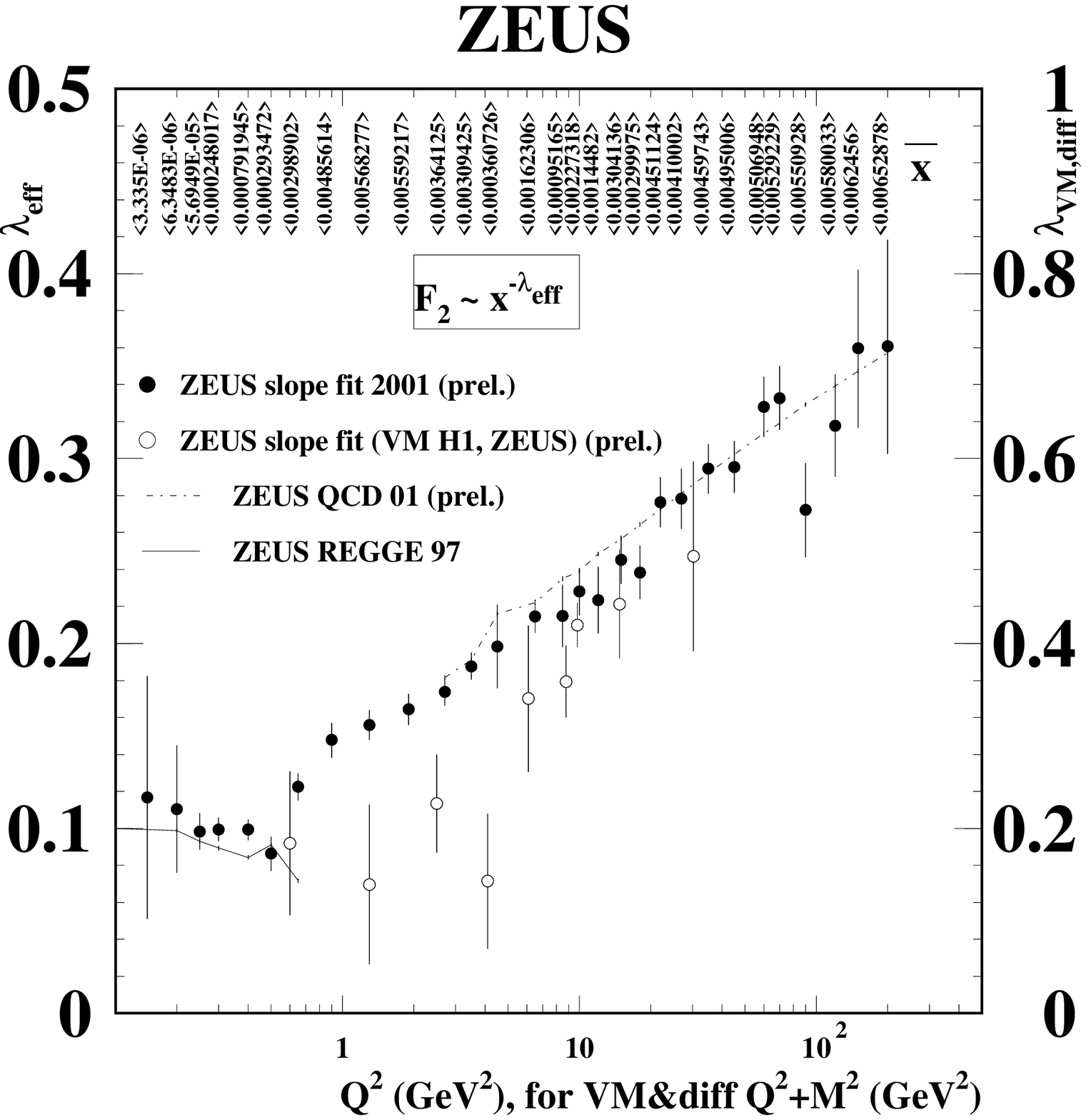,width=7.5cm}
\caption{$\lambda_V$ and  $\lambda_{eff}$ are compared (see
text).} \label{fig:lambda} }

It has been shown that the $W$ and $t$ behaviors of the elastic VM
production cross-sections do not depend strongly on the VM wave
functions; the variable $Q^2+M^2_V$  is a good choice of the scale
of this
interaction~\cite{hep-ph_99_08519,naroska,hep-ph_00_11050}. The
relative production rate for different VM's may depend on the
details of the wave functions. Data seem to indicate that the
production rate of heavy VM's is enhanced with respect to the
light VM's~\cite{hep-ph_00_11050,eps01_addendum} after taking into
account the SU(4) ratios.

The parameters of the Pomeron trajectory, have been extracted from
a number of different VM's for different photon virtualities. The
values of $\alpha_{\pom}(0)$ show a clear dependence with
$Q^2+M^2_V$. More data are needed to establish a clear correlation
between $\alpha_{\pom}^{\prime}$ and $Q^2+M^2_V$. The data do not
accommodate a universal Pomeron trajectory with fixed parameters.


Figure~\ref{fig:lambda} shows the values of $\lambda_V$ ($\left.{d
\sigma(\gamma p\rightarrow V p) \over dt }\right|_{t=0}$ is fitted
to $A x^{-\lambda_V}$) obtained for various VM's as a function of
$Q^2+M^2_V.$  $\lambda_{V}$ is compared with $\lambda_{eff}$
obtained from the DIS total cross-section~\footnote{Recent results
of $F_2$ released by the ZEUS Collaboration~\cite{epj_21_3} have
been used for this fits~\cite{eps01_surrow}.} ($F_2\propto
x^{-\lambda_{eff}}$ at fixed $Q^2$). The behavior of the steepness
of the rise of the VM production cross-section as $x\rightarrow 0$
with changing $Q^2+M^2_V$ is similar to the $x$ dependence of
$F_2$ with changing $Q^2.$ At small values of the scale ($Q^2$ and
$Q^2+M_V^2$ for DIS total and VM cross-sections, respectively)
$\lambda_{eff}$ and $\lambda_{V}$ depend weakly on the scale. This
correlation becomes stronger at larger scales. The similarities
between $\lambda_{V}$ and $\lambda_{eff}$ are evident. This
qualitative discussion serves as an indication that the dynamics
that drive the increase of the DIS total and VM cross-sections are
very similar. From the point of view of the dipole model, both
cross-sections may be easily related to each
other~\cite{hep-ph_0101085}.  A theoretical description of
$\sigma_{q\overline{q}}$ should be able to describe simultaneously
both the DIS total and VM cross-sections. The addition of VM data
to the analysis of $\sigma_{q\overline{q}}$ enhances the
sensitivity of HERA data to issues like the interface between hard
and soft physics and the sensitivity to saturation
effects~\cite{eps01_kow_gots}.

\providecommand{\href}[2]{#2}\begingroup\raggedright\endgroup


\end{document}